\documentclass{ws-ijmpcs}
\usepackage{color}
\usepackage{bm}
\usepackage{graphicx}
\usepackage{comment}
\usepackage{epsfig}
\usepackage{latexsym} 
\usepackage{multirow}
\usepackage{slashed}
\usepackage{rotating} 
\usepackage{dcolumn}
\usepackage{url,overcite}
\newcommand{\beq}{\begin{eqnarray}}
\newcommand{\eeq}{\end{eqnarray}}
\newcommand{\be}{\begin{equation}}
\newcommand{\ee}{\end{equation}}
\newcommand{\bw}{\begin{widetext}}
\newcommand{\ew}{\end{widetext}}
\newcommand{\ba}{\begin{array}}
\newcommand{\ea}{\end{array}}

\newcommand{\calC}{\mathcal{C}}
\newcommand{\calD}{\mathcal{D}}
\newcommand{\calM}{\mathcal{M}}
\newcommand{\calO}{\mathcal{O}}
\newcommand{\calZ}{\mathcal{Z}}
\begin{document}

\markboth{C. Liu}
{Recent results from Lattice QCD}

%
\catchline{}{}{}{}{}
%

\title{RECENT RESULTS FROM LATTICE QCD%
}

\author{CHUAN LIU%
}

\address{School of Physics and Center for High Energy Physcis\\
Peking University, Beijing 100871, China\\%
liuchuan@pku.edu.cn}



\maketitle

\begin{history}
\received{Day Month Year}
\revised{Day Month Year}
\end{history}

\begin{abstract}
 Recent Lattice QCD results  are reviewed with an emphasis on spectroscopic
 results concerning the charm quark. It is demonstrated that, with accurate computations from
 lattice QCD in recent years that can be compared with the existing or upcoming experiments,
 stringent test of the Standard Model can be performed which will greatly
 sharpen our knowledge on the strong interaction.
 \keywords{lattice QCD, charm quark, comparison with experiments.}
\end{abstract}

\ccode{PACS numbers:12.38.Gc, 11.15.Ha}

\section{Introduction}	
 In the past ten years or so, considerable progress has been achieved in
 lattice Chromodynamics (lattice QCD). Here, I will try to review briefly
 some selected results, with an emphasis on those related to the charm quark,
 and compare them with the experiments so far or possibly in
 the near future. For general recent lattice results, please consult
 the Lattice 2013 website~\cite{lattice2013webpage} where the talks are
 available online.

 Lattice QCD is a non-perturbative theoretical method that relies
 on Monte Carlo estimation of  physical
 quantities using gauge field samples that are generated
 according to a lattice action. Given a lattice action
 $S=S_g[U_\mu]+S_f[\bar{\psi},\psi,U_\mu]=S_g[U_\mu]+\bar{\psi}_x\calM_{xy}[U_\mu]\psi_y$,
 where $\calM[U_\mu]$ is called the fermion matrix
 \footnote{The explicit form of $\calM[U_\mu]$ depends on the type of lattice fermion (staggered, Wilson, etc.) used.},
 a physical quantity of interest, $\calO[\bar{\psi},\psi,U_\mu]$, which is built from the
 basic fields is given by the {\em ensemble average}:
 \be
 \label{eq:ensemble-average}
 \langle\calO\rangle={1\over\calZ}\int\calD\bar{\psi}\calD\psi\calD U_\mu
 \calO[\bar{\psi},\psi,U_\mu]e^{-S[\bar{\psi},\psi,U_\mu]}\;.
 \ee
 Here the partition function $\calZ$ is given by the relevant path integral
 \be
 \label{eq:partition-function}
 \calZ=\int\calD\bar{\psi}\calD\psi\calD U_\mu
 e^{-S[\bar{\psi},\psi,U_\mu]}=\int\calD U_\mu
 e^{-S_g[U_\mu]}\det\calM[U_\mu]\;.
 \ee
 Following the above equations, a typical lattice calculation therefore consists of two steps:
 In the first step, also known as the generation step, one generates the gauge field
 configurations according to the probability distribution: $P[U_\mu]=\calZ^{-1}e^{-S_g[U_\mu]}\det\calM[U_\mu]$
 and stores them for later usage; In the second step, also known as the measurement step,
 any interested observable $\calO[\bar{\psi},\psi,U_\mu]$ is measured from the pre-stored gauge field
 configurations with the quark and anti-quark fields in the corresponding observable
 replaced by the corresponding quark propagators, which are relevant matrix elements of $\calM^{-1}[U_\mu]$,
 in a particular gauge field background. Note that the fermion matrix $\calM[U_\mu]$ is diagonal
 in flavor space. Therefore, the determinant of the matrix in Eq.~(\ref{eq:partition-function})
 is in fact a product of determinants for each quark flavor.
 Depending on how many flavors are kept in the generation step,
 we call them $N_f$ flavor lattice QCD.
 \footnote{Quenched lattice QCD corresponds to $N_f=0$ in this sense.}

\section{Spectrum calculations}

\subsection{Hadron Masses, conventional computation}

 To compute the mass values of hadrons, one starts from a set of
 interpolating operators $\{\calO_\alpha(t):\alpha=1,2,\cdots,N\}$~\cite{lattice2013Thomas}.
 These operators carry the correct quantum numbers such that
 the state $\calO^\dagger_\alpha(t)|\Omega\rangle$, with $|\Omega\rangle$ being
 the QCD vacuum state,  has the same
 quantum numbers as those of the hadron in question.
 In lattice Monte Carlo simulations, the following correlation matrix is measured:
 \be
 \calC_{\alpha\beta}(t)=\langle\Omega|\calO_\alpha(t)\calO^\dagger_\beta(0)|\Omega\rangle
 \;.
 \ee
 On the other hand, it is known that the same correlation matrix is given by
 \be
 \calC_{\alpha\beta}(t)=\sum_n{e^{-E_n t}\over 2E_n}Z^{(n)}_\alpha Z^{(n)*}_\beta
 \;,
 \ee
 where the summation is over all eigenstates (labelled by $n$) of the Hamiltonian with
 the $E_n$'s being the corresponding eigenvalues.
 The measured correlation matrix $\calC_{\alpha\beta}(t)$ is then
 passed through a standard variational calculation, also known as
 the generalized eigenvalue problem (GEVP)~\cite{Luscher:1990ck,Dudek:2010wm},
 for some given $t_0$:
  \be
 \calC_{\alpha\beta}(t)v^{(n)}_\beta=\lambda^{(n)}(t)\calC_{\alpha\beta}(t_0)v^{(n)}_\beta
 \;.
 \ee
 Here $\lambda^{(n)}(t)$'s yield the eigenvalues of the Hamiltonian,
 namely the $E_n$'s via
 \be
 \lambda^{(n)}(t)\sim e^{-E_n(t-t_0)}\left[
 1+O\left(e^{-\Delta E(t-t_0)}\right)
 \right]\;,
 \ee
 while the eigenvectors $v^{(n)}_\beta$ are related to the corresponding
 overlap $Z^{(n)}_\alpha$~\cite{Dudek:2010wm}.

 Standard light hadron spectroscopy has been studied by various groups
 in recent years, see e.g. Ref.~\refcite{Durr:2008zz}.
  \begin{figure}[htb]
 \includegraphics[width=0.85\textwidth]{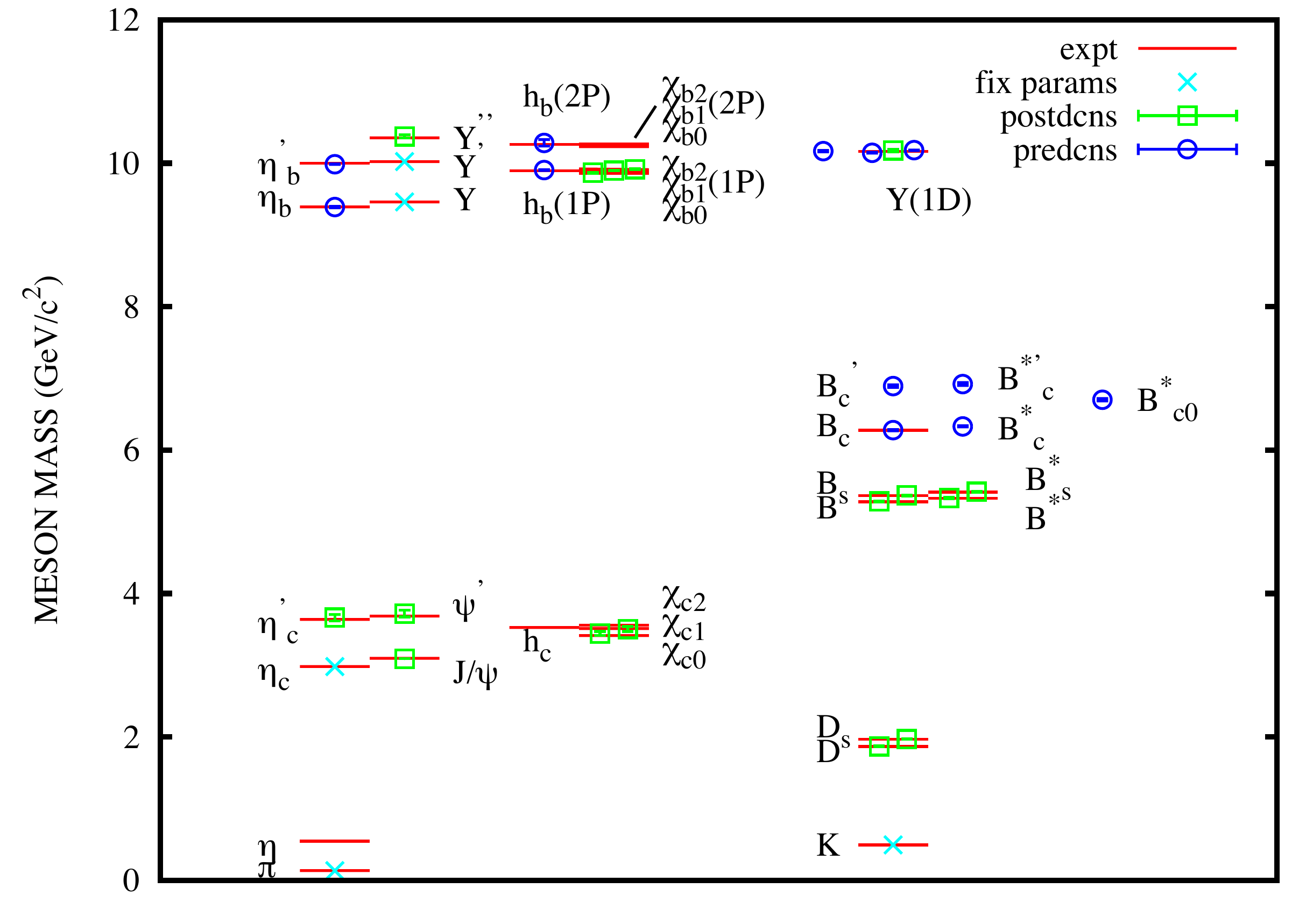}
 \caption{Heavy meson spectrum obtained from $N_f=2+1+1$ lattice QCD,
 taken from the paper by HPQCD collaboration~\cite{McNeile:2012qf}. \label{fig:heavymesons}}
 \end{figure}
 Hadrons containing the heavy quarks have also been studied.
 In Fig.~\ref{fig:heavymesons}, we have shown the heavy meson spectrum from HPQCD collaboration
 obtained using $N_f=2+1+1$ staggered quark configurations~\cite{McNeile:2012qf}.
 It is seen that, both post-dictions and predictions
 agree with the experiments, where available, astonishingly well.

\subsection{Hadron Masses, multi-hadron scattering effects}

 However, one should keep in mind that, in principle, these $E_n$'s are {\em not} the
 mass values of the hadrons. The eigenvalue of the QCD Hamiltonian in a particular symmetry
 sector is only approximately equal to the mass of the hadron if the hadron being studied is
 a narrow resonance within strong interaction.
 \footnote{If the hadron is stable in QCD, then the $E_n$ {\em is indeed}
 the mass of the hadron. But for resonances, which is true for majority
 of the hadrons, this is not the case.}
 Therefore, to really study a genuine hadronic resonance, one should
 study the scattering of hadrons. In fact, L\"uscher has established a formalism~\cite{Luscher:1990ux}
 in which the eigenvalues of the finite-volume Hamiltonian are
 related to the scattering phases of the two particles~\cite{lattice2013doring}.

 In the simplest case (single channel, s-wave scattering, neglecting higher $l$ contributions, etc.),
 this relation reads:
 \be
 \tan\delta_0(E)={\pi^{3/2}q\over \calZ_{00}(1,q^2)}
 \;\;,
 \ee
 where $\calZ_{00}(1,q^2)$ is the zeta function that can be computed accurately
 and $q$ is related to $E$ via
 \be
 E=\sqrt{m^2_1+\bar{k}^2}+\sqrt{m^2_2+\bar{k}^2}\;,
 \;\; q\equiv {\bar{k}L\over 2\pi}\;.
 \ee
 The lattice calculation goes the same way as we described above with the exception that the interpolation
 operators being used, namely the operators $\{\calO_\alpha(t):\alpha=1,2,\cdots,N\}$,
 should also include the two-particle operators.
 Using L\"uscher's formalism, one simply obtains the energy eigenvalues $E_n$'s which are then substituted into
 L\"uscher formula for $\delta_0(E_n)$.
 For small momentum close to the threshold, one could use the effective range expansion
 \be
 k\cot\delta(E)={1\over a_0}+{1\over 2}r_0k^2+\cdots
 \ee
 with $a_0$ being the scattering length and $r_0$ the effective range.

 \begin{figure}[t]
 \includegraphics[width=0.85\textwidth]{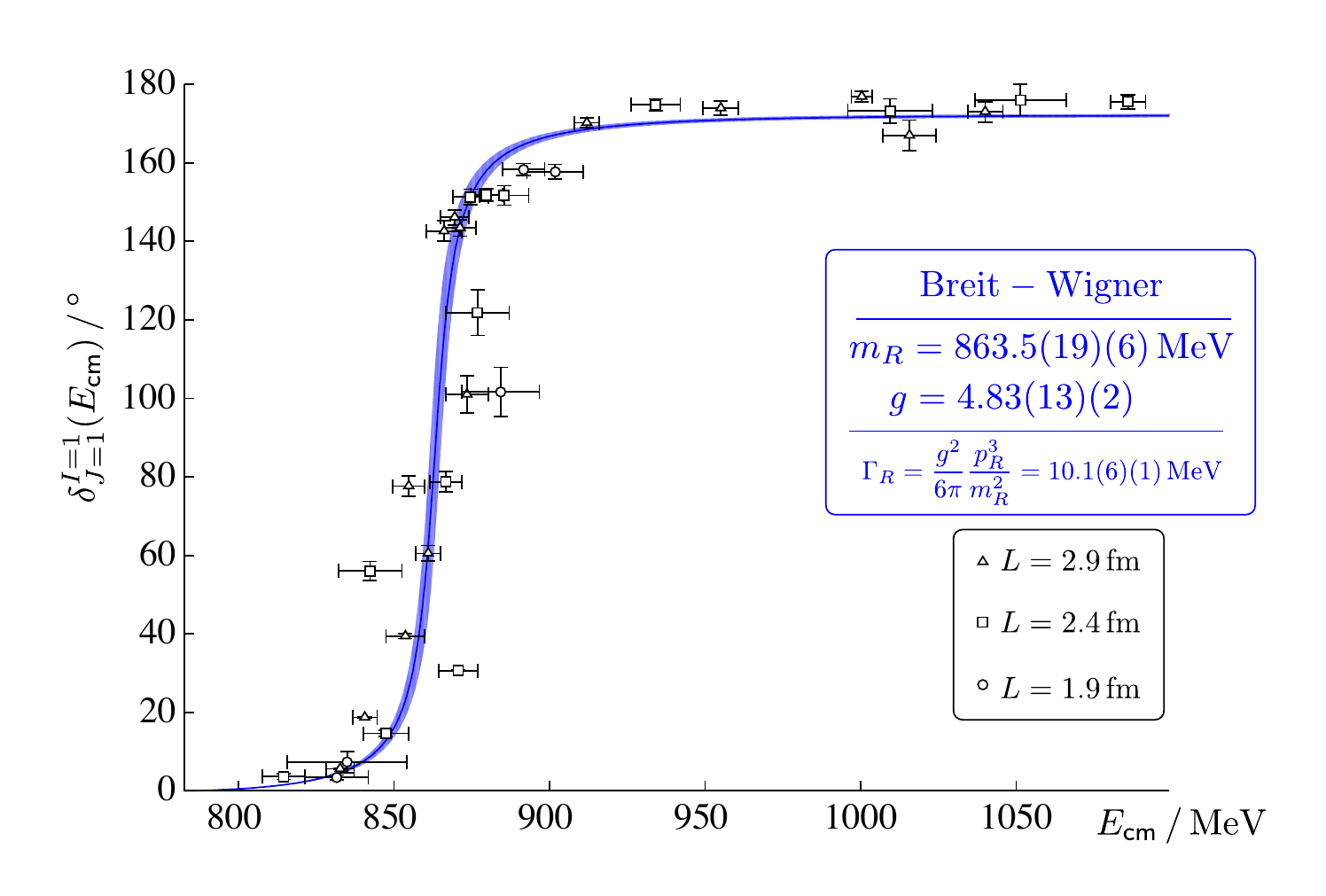}
 \caption{Phase shifts of $\pi\pi$ scattering in the $I=J=1$ channel
 calculated from lattice QCD using L\"uscher's formalism, taken
 from a paper by the Hadron Spectrum Collaboration~\cite{Dudek:2012xn}. \label{fig:delta}}
 \end{figure}
 L\"uscher's formula has been generalized to various cases, see e.g. Ref.~\refcite{Li:2012bi}
 and references therein.
 It has also been utilized successfully in lattice studies of
 hadronic resonances in recent years, see e.g. Ref.~\refcite{lattice2013Thomas,lattice2013doring}
 and references therein.
 A very good example is the lattice QCD study of the $\rho$ meson, a
 typical hadronic resonance in the $I=J=1$ channel of $\pi\pi$ scattering.
 In Fig~\ref{fig:delta}, taken from Ref.~\refcite{Dudek:2012xn}, we have shown the phase shifts of $\pi\pi$
 scattering in the $I=1$, $J=1$ channel computed within L\"uscher's formalism.

\subsection{Scattering of Charmed mesons}

 Recently, there have been numerous new hadronic structures observed
 which contain charm and anti-charm quark. These have been termed
 the $XYZ$ particles, see Refs~\refcite{Choi:2003ue,Aubert:2005rm,Choi:2007wga}.
 What is noticeable is that many of these
 newly observed states are close to the threshold of two
 known charmed mesons. For example, the $X(3872)$ is close
 to the threshold of $D$ and $D^*$ and so is the newly
 discovered $Z^\pm_c(3900)$~\cite{Ablikim:2013mio}.

 Several years ago, CLQCD has studied the $Z(4430)$ state observed
 by BELLE collaboration in quenched lattice QCD using L\"uscher formalism~\cite{Meng:2009qt}.
 Asymmetric volumes were used to investigate the low-momentum behavior
 of the scattering phase close to the threshold of $D^*$ and a $\bar{D}_1$.
 The investigation was done in the channel of $J^P=0^-$ and
 the interaction between the $D^*$ and a $\bar{D}_1$ was
 found to be attractive but not strong enough to form a genuine bound state.

 Recently, Prelovsek {\it et al} have studied the case of $X(3872)$
 using $N_f=2$ improved Wilson fermion configurations~\cite{Prelovsek:2013cra}.
 They claim that they have found some evidence for the state.
 With the same set of configurations, they have also studied the
 case of $Z_c(3900)$ with no signal of the state~\cite{Prelovsek:2013xba}.

 In summary, the study for these $XYZ$ states are still an on-going project.
 Lattice results obtained so far are still not systematic (most of them
 are at one lattice spacing, one volume, one pion mass etc.) and therefore
 one still cannot draw definite conclusions yet.
 The situation is still somewhat murky and more work needs to be done in
 the future to clarify the nature of these states from lattice QCD.

\subsection{Decay constants and the story of $f_{D_s}$}

  \begin{figure}[htb]
 \includegraphics[width=0.65\textwidth]{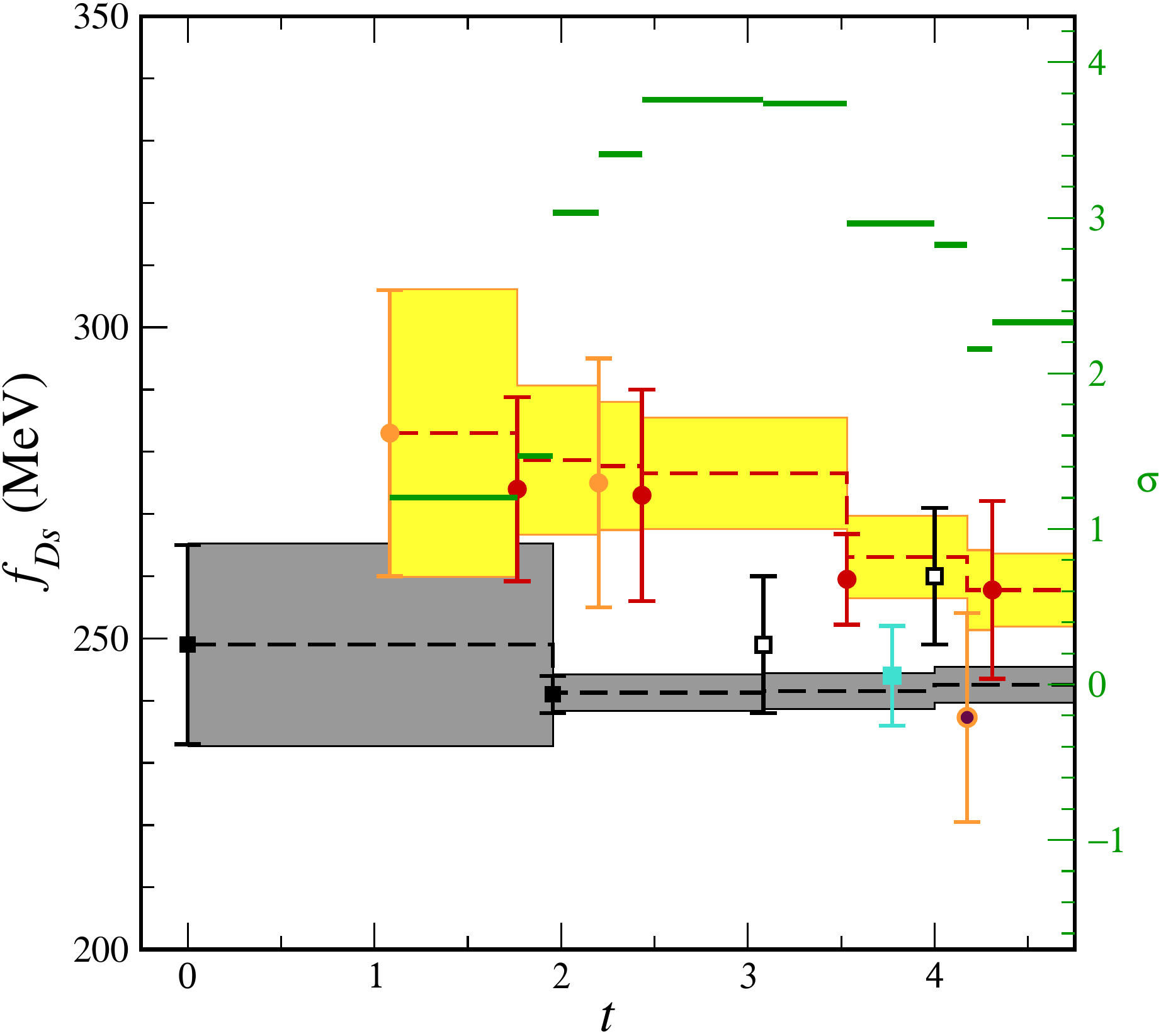}
 \caption{The history of $f_{D_s}$ till the end of 2009, taken
 from A.~Kronfeld's review talk ``The $f_{D_s}$ puzzle'' for PIC2009. \label{fig:fDs}}
 \end{figure}
 For a pseudoscalar meson, the decay constant is defined via the
 matrix elements
 \be
   \left\{ \begin{aligned}
 \langle\Omega|\bar{s}\gamma_\mu\gamma_5 c|D_s(p)\rangle &=if_{D_s}p_\mu\;,
 \\
 (m_c+m_s)\langle\Omega|\bar{s}\gamma_5 c|D_s(p)\rangle &=-m^2_{D_s}f_{D_s}
 \end{aligned}\right.
 \ee
 with the two definitions related to each other by PCAC.

 This quantity can be computed accurately in lattice QCD which then
 can be compared with the experiments. Around the year of 2008, some
 puzzling effects occurred that created a $3.8\sigma$ difference
 in the comparison of the lattice results and the experiments.
 For a more detailed description about this dilemma, the reader
 is referred to A.~Kronfeld's review talk ``The $f_{D_s}$ puzzle''
 in the proceedings of Physics In Collisions in 2009 (PIC2009)~\cite{Kronfeld:2009cf}.

 Basically, the puzzle came about because around 2008, the error of the lattice computation
 dropped significantly due to the HPQCD's new result on $f_{D_s}$~\cite{Follana:2007uv}.
 The tension with the experiments then increased to about $3.8\sigma$ around that time, making people
 contemplating about possible new physics to clarify the puzzle. However, as time goes
 on, this tension is finally eased, mainly due to the fact that
 new experimental results~\cite{HFAGwebpage}
 came down quite a bit while the lattice result also moves up a little,
 so that by the end of 2010, the discrepancy is only $1.6\sigma$~\cite{Davies:2010ip}.
 Most recent lattice calculations yields compatible results with those in 2009, with the errors
 further reduced, see e.g. C. Bernard's talk at Lattice 2013~\cite{lattice2013bernard}.
 Note that, although there is no significant ``puzzle'' right now for
 $f_{D_s}$, the story of the the so-called ``$f_{D_s}$ puzzle'' has given us
 a very instructive lesson: precise comparison between the experiments and theory is vital in this game.
 In the future, should there be more accurate experimental results
 come about, e.g. at BESIII, there could be further puzzles and
 by resolving these puzzles, we could sharpen our knowledge about QCD and beyond.

\section{Radiative transitions and decays of charmonia}

 Charmonium states play an important role in our understanding of QCD. For charmonia lying below
 the open charm threshold, radiative decays are important to illuminate
 the structure of these states. Recently and in the years to come,
 more experimental data are being
 accumulated at BEPCII which will enable us to make precise comparison
 between the experiments and theory.

 To lowest order in QED, the amplitude for $J/\psi\rightarrow \gamma H$ is given by
 \begin{equation}
 M_{r,r_\gamma,r_H}=\epsilon_{\mu}^*(\vec{q},r_\gamma)
 \langle H(\vec{p}_f,r_H)|j^{\mu}(0)|J/\psi(\vec{p}_i,r)\rangle,
 \end{equation}
 where $\vec{p}_i$/$\vec{p}_f$ is the initial/final three-momentum of
 the hadron, respectively, while $\vec{q}=\vec{p}_i-\vec{p}_f$ is
 the three-momentum of the real photon. We use $r$, $r_H$, $r_\gamma$ to denote
 the polarizations of the relevant particles.
 \footnote{If the final hadron $H$ is a scalar, then the label $r_H$ is not needed.}
 $\epsilon_\mu(\vec{q},r_\gamma)$ is the polarization
 four-vector of the real photon and $j^\mu_{\rm e.m.}(0)$ stands for the
 electromagnetic current operator due to the quarks.
 We emphasize that the hadronic matrix element
 $\langle H(\vec{p}_f,r_H)|j^{\mu}(0)|J/\psi(\vec{p}_i,r)\rangle$ is
 non-perturbative in nature and consequently should be computed
 using non-perturbative methods like lattice QCD.

 It turns out that the above mentioned matrix element can be obtained from the following
 three-point correlation function,
 \be
 \label{eq:3point}
 \Gamma_{i,\mu,j}^{(3)}(\vec{p}_f,\vec{q};t_f,t) =
 \frac{1}{T}\sum\limits_{\vec{y},\tau=0}^{T-1}
 e^{-i\vec{q}\cdot \vec{y}} \langle
 \Phi^{(i)}(\vec{p}_f,t_f+\tau)J_\mu (\vec{y},t+\tau)
 O_{V,j}(\vec{0},\tau)\rangle\;,
 \ee
 where $J_\mu=\bar{c}\gamma_\mu c$ is the vector current
 of the charm quark. By measuring three-point function in
 Eq.~(\ref{eq:3point}) together with relevant two-point functions,
 one could obtain the desired hadronic matrix element
 $\langle H(\vec{p}_f,r_H)|j^{\mu}(0)|J/\psi(\vec{p}_i,r)\rangle$.

 For charmonium states various lattice computations, both quenched~\cite{Dudek:2006ej}
 and unquenched~\cite{Chen:2011kpa,Donald:2012ga,Becirevic:2012dc},
 have been performed and the results can be compared with the corresponding experiments.
 For example, for the radiative transition rate $J/\psi\rightarrow\gamma\eta_c$, it
  is parameterized by a form factor $V(q^2)$:
 \beq
\langle\eta_c(p')|\bar{c}\gamma^\mu c|J/\psi(p)\rangle
={2V(q^2)\over M_{J/\psi}+M_{\eta_c}}\epsilon^{\mu\alpha\beta\gamma}
p'_\alpha p_\beta\epsilon(p)_\gamma\;,
 \eeq
 and the decay rate is given by the value of $V(0)$.
 The HPQCD result finally gives:~\cite{Donald:2012ga}
 \beq
 V(0)=1.90(7)(1)\;,
 \eeq
 which is larger than the experimental value from CLEO~\cite{Mitchell:2008aa} by about $1.7\sigma$.
 Another lattice calculation using $2$ flavors of twisted mass fermions
 yields even larger (but compatible with that of HPQCD) result~\cite{Becirevic:2012dc}.
 Right now, the experimental error is larger than those in lattice computations.
 Later on, more accurate experiments at BESIII will surely bring more stringent
 test to this comparison.

 Not only can one computes transitions among charmonium states, one could
 also compute the radiative transitions of $J/\psi$ to pure-gauge glueballs on
 the lattice. This has been done recently by CLQCD in quenched lattice
 QCD~\cite{Gui:2012gx,Yang:2013xba}.
 The relevant hadronic matrix element $\langle G(\vec{p}_f, r_G) | J_{\mu}(0) |V(\vec{p}_i,r)\rangle$
 are expanded in terms of form factors and known kinematic functions, see e.g. Ref.~\refcite{Yang:2012mya}.
 Take the tensor glueball as an example, we have
  \begin{eqnarray}
 \langle G(\vec{p}_f, r_G) | J_{\mu}(0) |V(\vec{p}_i,r)\rangle &=& \alpha_1^\mu E_1(Q^2) + \alpha_2^{\mu}M_2(Q^2)\nonumber \\
 &+& \alpha_3^\mu E_3(Q^2) + \alpha_4^\mu C_1(Q^2)+ \alpha_5^\mu C_2(Q^2),
 \end{eqnarray}
 where $\alpha^\mu_i$'s are kinematic functions and $E_1$, $M_1$, $E_2$ etc. are
 the form factors. For the scalar glueball, one only needs two form
 factors $E_1(Q^2)$ and $C_1(Q^2)$ instead.

   \begin{figure}[htb]
 \includegraphics[width=0.95\textwidth]{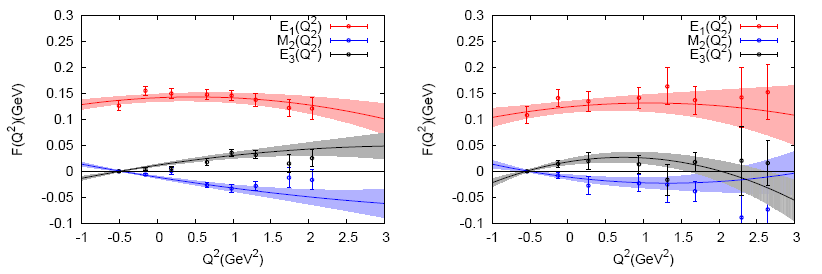}
 \caption{The extracted form factors for tensor glueball at two different
 lattice spacings~\cite{Yang:2013xba}. \label{fig:glueball}}
 \end{figure}
 The physical decay rates for $J/\psi\rightarrow\gamma G$ only depend on the
 values of the form factors at $Q^2=0$:
   \begin{eqnarray}
  \label{eq:scalar_width}
\Gamma(J/\psi\rightarrow \gamma
G_{0^{++}})&=&\frac{4\alpha|\vec{p}_\gamma|}{27M_{J/\psi}^2}|E_1(0)|^2,
 \\
 \label{eq:tensor_width}
 \Gamma(J/\psi\rightarrow \gamma
G_{2^{++}})&=&\frac{4\alpha|\vec{p}_\gamma|}{27M_{J/\psi}^2}\left(|E_1(0)|^2+|M_2(0)|^2+|E_3(0)|^3\right).
 \end{eqnarray}
 Therefore, the computed matrix elements at various values of $Q^2$ are then fitted
 using a polynomial in $Q^2$ to extract the relevant values at $Q^2=0$.
 In Fig.~\ref{fig:glueball}, we show the case for the tensor glueball
 at two different lattice spacings. After the continuum extrapolation,
 the physical values for these form factors are obtained which then
 gives us the prediction of relevant decay rates.

 Needless to say, the so-called pure-gauge glueballs are not physical objects
 that can be measured directly in experiments. In real world they mix with
 ordinary hadrons with the same quantum numbers. However, the above mentioned
 lattice calculation can provide us with important information about the pure-gauge glueball
 component in the measured hadronic states. For example, this calculation helps to clarify,
 say in the scalar channel, which of the three candidates $f_0(1370)$, $f_0(1500)$
 and $f_0(1710)$ contains more glueball component and can thus be regarded as
 the best candidate for the scalar glueball~\cite{Gui:2012gx}.

\section{The anomalous magnetic moment of the muon}

 A precisely measured quantity that brings significant deviation from
 the prediction of the Standard Model (SM) is the anomalous magnetic moment
 of the muon. The definition of this quantity is
 \be
 a_\mu=(g_\mu -2)/2\;.
 \ee
 This is one of the most accurately  measured quantities:
 \be
 a^{\rm exp}_\mu=11659208.9(5.4)(3.3)\times 10^{-10}\;.
 \ee
 The same quantity predicted by the Standard Model is, however,
 \be
 a^{\rm SM}_\mu =11659180.2(0.2)(4.2)(2.6)\times 10^{-10}.
 \ee
 The difference of the above two equations is
 \be
 \Delta a_\mu\equiv a^{\rm exp}_\mu-a^{\rm SM}_\mu =287(63)(49)\times 10^{-11}.
 \ee
 This is roughly a $3.6\sigma$ effect, one of the remaining ``puzzles'' in SM.

 Although the major part of $a_\mu$ comes from QED, the major {\em theoretical uncertainty}
 comes from the hadronic contributions, denoted as $a^{\rm Had}_\mu$, and is given by
 \be
 a^{\rm Had}_\mu=a^{\rm HVP}_\mu + a^{\rm HLbL}_\mu+\cdots\;.
 \ee
 Here, the leading contribution comes from
 Hadronic Vacuum Polarisation (HVP), $a^{\rm HVP}_\mu$,
 that can be measured experimentally with the help of dispersion relations.
 The next-order correction, the Hadronic Light-by-light scattering (HLbL), $a^{\rm HLbL}_\mu$,
 cannot be related to experimentally measurable quantities and currently
 relies on modelling, see e.g. PDG reviews~\cite{Beringer:1900zz}.

 There have been several lattice attempts to compute $a^{\rm HVP}_\mu$ since past
 decade~\cite{Blum:2002ii,Gockeler:2003cw,Aubin:2006xv,Feng:2011zk,Boyle:2011hu,DellaMorte:2011aa},
 which can be related to the current-current correlator in QCD:
 \be
 a^{\rm HVP}_\mu=\alpha^2\int^\infty_0 {dQ^2\over Q^2} w(Q^2/m^2_\mu)\hat{\Pi}(Q^2)
 \;,
 \ee
 where $w(Q^2/m^2_\mu)$ is a known function~\cite{Blum:2002ii};
 The quantity $\hat{\Pi}(Q^2)=\Pi(Q^2)-\Pi(0)$ is defined via:
 \be
   \left\{ \begin{aligned}
 \Pi_{\mu\nu}(Q) &\equiv (Q_\mu Q_\nu-Q^2\delta_{\mu\nu})\Pi(Q^2)\\
 \Pi_{\mu\nu}(Q) &=\int d^4x e^{iQ\cdot x}
 \langle\Omega|T[J_\mu(x)J_\nu(0)]|\Omega\rangle
 \end{aligned} \right.\;.
 \ee
 In principle, the  current-current correlator $\langle\Omega|T[J_\mu(x)J_\nu(0)]|\Omega\rangle$
 can be computed using standard methods in lattice QCD. However, it does pose several
 challenges for the current state of the art lattice computations.
 It turns out that this quantity is dominated by contributions
 at low $Q^2$, typically $Q^2\sim 1$GeV$^2$, which is difficult for
 current realistic lattice calculations. It also requires rather
 delicate error controlling in lattice extrapolations in order to
 match the precision of the experimental measurements.
 Right now, the lattice results are in no comparison with
 those using dispersion relations yet, as far as the errors are concerned.
 However, with more and more lattice groups are joining the game,
 see e.g. Lattice 2013 talks~\cite{lattice2013amutalks},
 hopefully we will get a  better control in the years to come.

 \begin{figure}[htb]
 \includegraphics[width=0.85\textwidth]{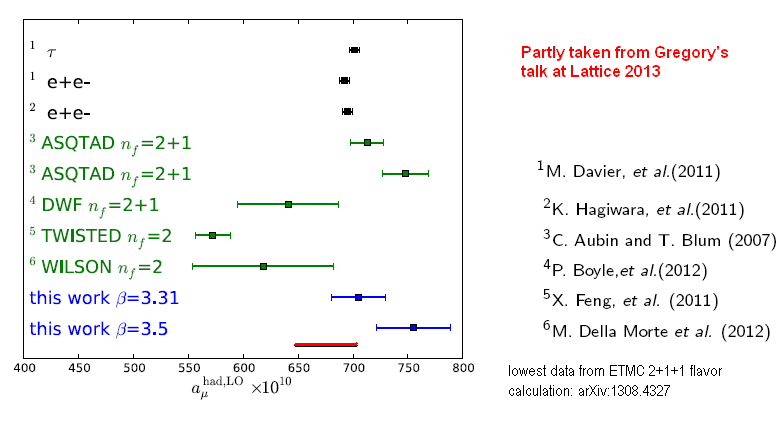}
 \caption{The current status of lattice computation for $a_\mu$,
 taken partly from Gregory's talk at Lattice 2013~\cite{lattice2013Gregory}. \label{fig:amu}}
 \end{figure}
 In Fig.~\ref{fig:amu},
 I show a summary figure taken partly from Gregory's talk~\cite{lattice2013Gregory}
 at Lattice 2013. I have added another new point from ETMC collaboration
 using $2+1+1$ twisted mass fermion configurations~\cite{Burger:2013jya}.
 All available lattice data for $a^{\rm HVP}_\mu\equiv a^{\rm had, LO}_\mu$
 are summarized together with those obtained using dispersion relations.
 It is seen that, although some of the lattice results are compatible with those of
  dispersive analysis, they still need much improvement
 to reduce the errors in order to match the accuracy for the dispersive
 analysis which is comparable to that of the experimental ones.

\section{Conclusions and outlook}

  In recent years, as a theoretical tool from first principles, lattice QCD has become
  an important player in relevant field of physics involving the
  strong interaction. Due to the crossover from quenched to unquenched
  calculations, more and more lattice results are available that
  are both practical and accurate enough to be compared with the existing
  or upcoming experiments. Some of these lattice results are reviewed
  here in this talk, with an emphasis on the properties involving the
  charm quark. I have also try to emphasize the interplay between the
  experimental results and those obtained from lattice computations.

  For the hadron masses and decays,
  we have seen both post-dictions and pre-dictions that agree rather well
  with the experiments. Lattice computations nowadays can also handle hadron-hadron scattering
  processes, which not only helps to obtain important information about hadronic interactions
  for the resonances, but also to clarify some of the
  newly discovered exotic hadronic states, the so-called $XYZ$ particles.
  For the radiative transition and decays of charmonium, lattice computations
  have matured to a stage that a detailed comparison with experiments is possible.
  It is also possible to offer us information about glueballs in the
  radiative decays.

  Many of the quantities mentioned above can be obtained rather precisely in lattice calculations,
  thus providing a precision test for QCD.
  It is important and instructive to compare the available lattice QCD results with the experiments.
  Some of the lattice results agree with the experiment perfectly: the hyperfine splitting of
  the charmonium $M_{J/\psi}-M_{\eta_c}$, for example;
  some even awaits further more accurate experiments: the charmonium radiative
  transition rate $\Gamma_{J/\psi\rightarrow \gamma\eta_c}$ and the decay constant $f_{D_s}$.
  There are also quantities that the lattice cannot compute accurately enough.
  An example is the muon $g-2$, a quantity showing $3.6\sigma$ deviation from the standard model.
  It is feasible to compute the leading hadronic contributions from
  QCD first principles, however, more efforts are needed to bring down the error bars
  in future lattice calculations.
  It is in this constant process of comparison between the experiments
  and theory that we could sharpen our understanding of the theory of strong interaction.

\section*{Acknowledgments}

 This work is supported in part by the
 National Science Foundation of China (NSFC) under the project
 No. 11335001 and No.11021092.
 It is also supported in part by the DFG and the NSFC (No.11261130311) through funds
 provided to the Sino-Germen CRC 110 ``Symmetries and the Emergence
 of Structure in QCD''.



\end{document}